\documentclass[twocolumn,prl,twoside,a4paper]{revtex4}

\usepackage{graphicx}
\usepackage[usenames,dvipsnames]{xcolor}
\usepackage{amsmath,bbm}

\begin{document}

\title{Split-sideband spectroscopy in slowly modulated optomechanics}

\author{E.~B. Aranas, P.~G.~Z. Fonseca, P. F. Barker}
\affiliation{Department of Physics and Astronomy, University College London, Gower Street, London WC1E 6BT, United Kingdom}
\author{T.~S.~Monteiro}
\email{t.monteiro@ucl.ac.uk}
\affiliation{Department of Physics and Astronomy, University College London, 
Gower Street, London WC1E 6BT, United Kingdom}

\begin{abstract}
Optomechanical coupling between the motion of a mechanical oscillator and a cavity represents a new arena
for experimental investigation of quantum effects on the mesoscopic and macroscopic scale.
 The motional sidebands of the output of a cavity offer ultra-sensitive probes of the dynamics.
We introduce a scheme whereby these sidebands split asymmetrically and show how they 
may be used as experimental diagnostics and signatures of quantum noise limited dynamics.
 We show split-sidebands with controllable asymmetry occur by simultaneously modulating the light-mechanical
 coupling $g$ and $\omega_M$  - slowly and  out of-phase. Such modulations are generic but already occur in
optically trapped set-ups where the equilibrium point of  the oscillator is varied cyclically.
 We analyse recently observed, but overlooked, experimental split-sideband asymmetries; although not yet in the 
quantum regime, the data suggests that split sideband structures are easily accessible to future experiments. 

\end{abstract}

\maketitle 

Cavity optomechanics offers rich possibilities for experimental investigation of 
the theory of quantum measurement and the role of quantum noise \cite{AKMreview,FMnoisereview}.
Several groups have successfully cooled a mechanical oscillator via its coupling to a mode of an electromagnetic cavity 
\cite{Teufel2011,Chan2011,Kipp2012}  down to its quantum ground state (or very close to it) i.e.  mean phonon occupancy $n_{ph} \lesssim 1$. Read-out of the temperature was achieved by detection of motional sidebands in the cavity output; the theory for
quantum sidebands was elucidated in \cite{Wilson2007,Marquardt2007} . The cavity fields serve a dual purpose: they provide not only 
 the laser cooling but also an ultrasensitive means for detection of displacements on the scale of quantum zero-point fluctuations; this has motivated considerable interest in quantum-limited measurements in this context,  following the early pioneering  work  by Braginsky and Khalili \cite{Braginsky}.

 An important development was the detection of an asymmetry \cite{SideAsymm2012,SideAsymm2012a,Weinstein} in the two frequency peaks (sidebands) of the output power of a probe beam detuned to the positive and negative side of the cavity resonance.  Albeit indirectly \cite{SideAsymm2012a,Weinstein}, the observations mirror an underlying asymmetry in the  motional spectrum: an oscillator in its ground state $n_{ph}=0$, can absorb a phonon and down-convert the photon frequency (Stokes process); but it can no longer emit any energy and up-convert a photon (anti-Stokes process). 

Sideband asymmetry has become  an important tool in optomechanics and has now been used to establish  cooling limited by only quantum backaction \cite{Peterson}.  Ponderomotive squeezing, whereby narrowband cavity output falls below the 
technical imprecision noise floor  is also of much current interest \cite{Safavi2013,Purdy2013,Pontin} though is also observed in  oscillators
in a high thermal state.

Recent rapid progress on cooling optically levitated systems suggests ground state cooling may be in sight \cite{Novotny,Ulbricht,arxiv2016}. This strongly motivates development of robust probes of 
the quantum dynamics. Such systems offer unique potential to sensitively probe quantum noises due to their near 
complete decoupling from environmental heating and decoherence. They also readily access the quantum shot-noise limit
\cite{Novotny}, since in a vacuum, the mechanical damping  $\Gamma_M\to 0$.

\begin{figure*}[ht]
{\includegraphics[width=6.9in]{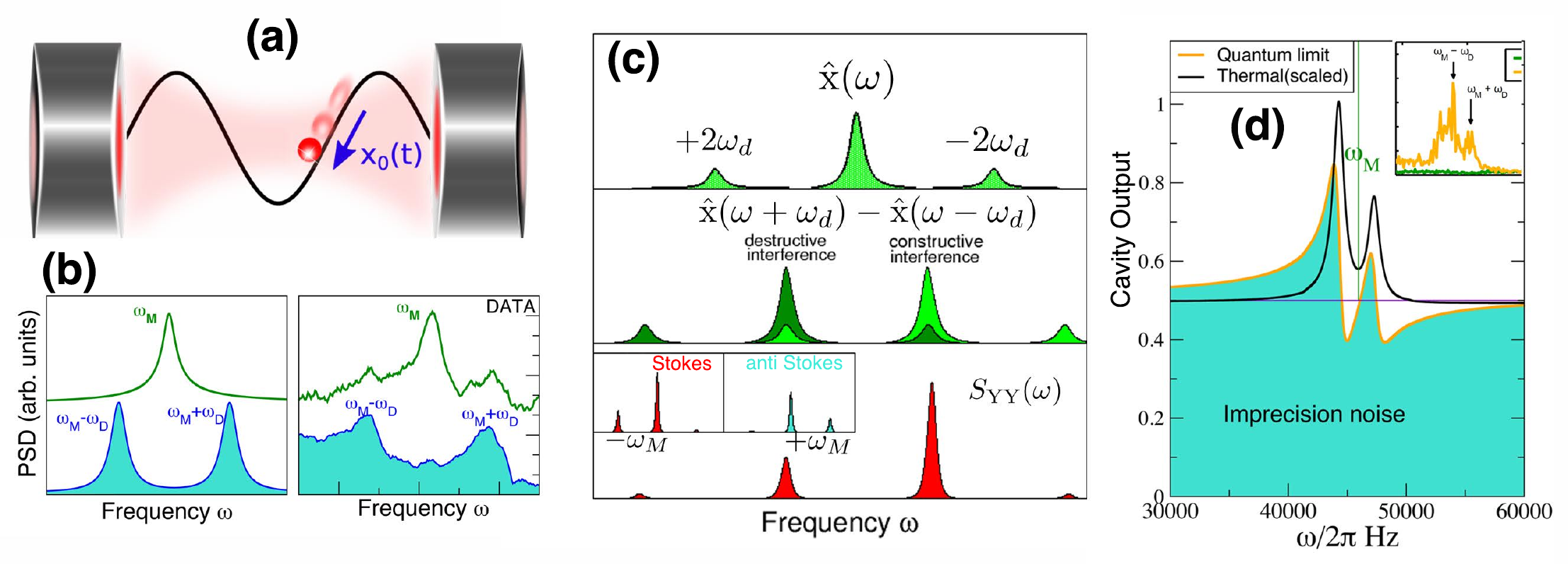}}
\caption{{\bf (a)} For set-ups where an oscillator is dipole-force trapped by the standing wave of a cavity mode there is no optomechanical cooling at
 the antinode $x_0=0$ (the potential minimum of an optical well of width $\lambda/2$), since there is
no light-matter coupling. Hence, such set-ups \cite{Kiesel2013,PRL2015} require an auxiliary field to pull the centre of the mechanical oscillation away from $x_0=0$. In electro-optical traps  \cite{PRL2015}, a slow oscillation is induced such
that $x_0(t)= \textrm{X}_d \sin{\omega_d t}$ . For small oscillations, we show  this corresponds to an effective
 modulation of the coupling coupling $g(t) =  2\overline{g} \sin{\omega_dt}$ and a simultaneous, out-of-phase,
modulation of the  mechanical frequency  $\omega_M(t)=\overline{\omega}_M + 2\omega_2 \cos{(2 \omega_dt)}$.
 {\bf (b)}  For a small ($\textrm{X}_d \ll \lambda$) modulation,   $ \omega_2 \approx 0$ and only $g$ is appreciably modulated.
In that case, while the displacement spectrum, $S_{\textrm{xx}}(\omega)\equiv \langle | {\hat{\textrm{x}}}(\omega)|^2\rangle$, 
 is still peaked at $\pm \omega \simeq \omega_M$, the experimental cavity spectrum,
($S_{\textrm{yy}}(\omega)$)  exhibits a characteristic structure of ``twin peaks''  at $\pm \omega =\omega_M\pm \omega_d$
\cite{arxiv2016}.
  {\bf (c)} For larger modulations,  the effect of $\omega_2 >0$ is to produce additional $\pm 2\omega_d$ side-peaks in
the ${\hat{\textrm{x}}}(\omega)$ spectrum. There is constructive enhancement of the $\omega_M-\omega_d$ 
peak, and destructive cancellation of the $\omega_M+\omega_d$ peak, so the ``twin peak'' structure is replaced by
a  pair of peaks of asymmetric heights. For small $\overline{g}$, the ratio between peaks $r \simeq (\omega_2-2\omega_d)^2/(\omega_2+2\omega_d)^2$
so the $\omega_M+\omega_d$ peak is strongly suppressed for $\omega_2 \sim 2\omega_d$ ($r\approx 0$). This asymmetry is distinct from the usual Stokes/antiStokes sideband asymmetry at 
$\pm \omega \simeq \omega_M$, which is still present.  {\bf (d)} In thermal regimes,  the ratio $r$ is insensitive to $\Gamma_M$;
 however, as $\Gamma_M \to 0$ and the backaction limit is attained,  correlations between back-action and incoming  noise alters the relative heights of the peaks, mainly since ponderomotive squeezing lowers the height of the $\omega_M+\omega_d$ peak relative to the imprecision floor.
 For incoming quantum shot noise,
significant  changes in $r$ arise only if the oscillator is near the ground state.
 Inset reproduces (classical) data from Fig.5 of \cite{PRL2015} which 
  supports our model of the split-sidebands.}
\label{Fig1}
\end{figure*}

A standard optomechanical system comprises a mechanical oscillator interacting with a
laser-driven cavity. In the frame rotating with the driving laser, typical experimental regimes using
 an extraordinarily broad range of physical platforms (cantilevers, microtoroids, membranes, photonic crystals) 
 are well described by the two-coupled oscillator Hamiltonian:

\begin{eqnarray}
{\hat{H}}/\hbar= \Delta {\hat{\textrm{a}}}^\dagger {\hat{\textrm{a} }} +
                 \omega_M ({\hat{\textrm{p}}}^2 +  {\hat{\textrm{x}}}^2) +
                 g ({\hat{\textrm{a}}}^\dagger+ {\hat{\textrm{a}}}) {\hat{\textrm{x}}}.
\label{OptoH}
\end{eqnarray}
  $\hat{\textup{a}}^\dagger$, $\hat{\textup{a}}$ are creation and annihilation operators
for small fluctuations cavity mode about its equilibrium value $\alpha$ while
 $\hat{\textup{x}} \equiv {\hat { \textrm b}} + {\hat{\textrm b}}^{\dag}$ (in appropriately scaled units) represents
 a small displacement of the mechanical oscillator about its equilibrium position $x_0$. 
Dissipative processes are treated by standard input-output theory, including input noises 
incident in the optical cavity  $\sqrt{\kappa}\hat{\textrm{a}}_{in}, \sqrt{\kappa}\hat{\textrm{a}}_{in}^\dag$
and mechanical oscillator $\sqrt{\Gamma_M}\hat{\textrm{b}}_{in}, \sqrt{\Gamma_M}\hat{\textrm{b}}_{in}^\dag$
where $\kappa,\Gamma_M$ are the cavity and mechanical damping rates while $g$ is the strength of the optomechanical coupling.

 However, here we consider instead a harmonically modulated optomechanical coupling
 $g(t) = 2\bar{g} \sin{\omega_d t}$ and 
mechanical frequency $\omega_M(t) = \overline{\omega}_M + 2{\omega}_2 \cos{2\omega_d t}$.
Other studies have investigated periodically modulated optomechanics, but interest has been focused on {\em  resonant }driving
$\omega_d \sim \omega_M, |\Delta|$ \cite{Mari2009,Malz} leading to  interesting effects like 
squeezing or  OMIT \cite{OMIT2010}.
In contrast, here we investigate systems which are modulated slowly $\overline{\omega}_M \gg  \omega_d$ (so as to preserve
linearisation about $x_0(t)$ and $\alpha$) and hence are far off-resonant. In addition, the $g,\omega_M$ modulations are  out of phase, in the sense that when the mechanical frequency is a maximum, the magnitude of coupling strength between motion and cavity field is a minimum; and vice-versa.

Added impetus to our study is provided by its relevance to optically trapped systems including levitated nanoparticles 
\cite{Lireview,Romero2010,Chang2010,Kiesel2013,Monteiro2013,Li2011,Gieseler2012}. 
We show here that the anti-phase $g,\omega_M$ modulation arises automatically if the mean position of the oscillator
varies harmonically $x_0(t) = X_D \sin{\omega_d t}$. We find it  accounts for previously unexplained asymmetries in the sidebands of levitated
oscillators, nanoparticles in hybrid electrical-optical  traps \cite{PRL2015,arxiv2016}. Such traps  to date provide the
only experimental realisation of stable trapping and  cooling of a nanoparticle at high vacuum, in a cavity.
When strongly cooled, the linearised dynamics of Eq.\ref{OptoH} dominate the dynamics \cite{arxiv2016}.
We emphasize the results of the study are generic to any set-up that can achieve these
  anti-phase $g,\omega_M$ modulations;  but we illustrate and test the model against 
numerics used to simulate hybrid trap dynamics \cite{arxiv2016} and  optically trapped particles, as illustrated in Fig.\ref{Fig1}(a).  

The  equations of motion  for the standard set-up Eq.\ref{OptoH} are solved \cite{AKMreview} in frequency space. In terms of quadrature operators $\hat{\textrm{y}}(\omega)=\frac{1}{\sqrt{2}}\left({\hat{\textrm{a}}}^\dagger(\omega)+ {\hat{\textrm{a}}}(\omega) \right)$
one may write:
\begin{equation}
\hat{\textrm{y}}(\omega) = ig \eta(\omega) \cdot  \hat{\textrm{x}}(\omega)  + \sqrt{\kappa} \hat{Y}_{th}(\omega)
\label{Yspec}
\end{equation}
where $\eta(\omega)=\chi_o(\omega)-\chi_o^*(-\omega)$ and $\hat{Y}_{th}(\omega)=\chi_o(\omega) \hat a_{in} + \chi_o^*(-\omega)\hat a_{in}^\dag$, while $\chi_o(\omega) = \left [ -i (\omega + \Delta) + \frac{\kappa}{2} \right ]^{-1}$
represents the optical susceptibility. In this well-known form, the first term represents the back-action of the mechanical motion on the cavity field, the second the cavity-filtered incoming quantum noise.  The measurable, cavity output spectrum is then obtained from input-output theory \cite{AKMreview}
${\hat{\textrm{a}}}_{out}(\omega)=\hat a_{in}-\sqrt{\kappa} {\hat{\textrm{a}}}(\omega)$ by considering the interference with the incoming
( imprecision noise, typically shot noise from the laser), so $\hat{\textrm{y}}_{out}(\omega)=\frac{1}{\sqrt{2}}[{\hat{\textrm{a}}}_{out}(\omega)+{\hat{\textrm{a}}}^\dagger_{out}(\omega)]$. 

If we  include the modulation of $g(t)$ we obtain instead:
\begin{equation}
\hat{\textrm{y}}(\omega) = \bar{g} \eta(\omega) \cdot  \left[\hat{\textrm{x}}(\omega+\omega_d)- \hat{\textrm{x}}(\omega-\omega_d)\right] + \sqrt{\kappa} {\hat Y}_{th}(\omega)
\label{Yshif}
\end{equation}
We elucidate details in \cite{SuppInfo}, but the notable difference between the standard case and the modulated optomechanics 
is that in Eq.\ref{Yshif} the optical field does not probe the displacement spectrum $\hat{\textrm{x}}(\omega)$
 but rather is sensitive to the interference between shifted spectra  at $\omega_M \pm \omega_d$.

For $\omega_2 \simeq 0$, the minus sign in Eq.\ref{Yshif} is not significant: the shifted spectra do not interfere appreciably.
The result is a cavity field fluctuation spectrum characterised by a ``twin peaks'' structure as illustrated in 
Fig.\ref{Fig1}(b)  and also in \cite{arxiv2016}. The green trace shows the phase of the Pound-Drever-Hall signal used to lock the cavity: while not a sensitive detection method, it follows the phase of the field so more directly represents $\hat{\textrm{x}}(\omega)$: the latter is peaked at $\omega=\omega_M$, in contrast with the cavity intensity modulations which are peaked at
$\omega_M \pm \omega_d$.

The effect of the frequency modulation, $\omega_M(t) = \overline{\omega}_M + 2{\omega}_2 \cos{2\omega_d t}$ however,
is to couple $\hat{\textrm{x}}(\omega)$ directly to $\hat{\textrm{x}}(\omega\pm 2\omega_d)$; in that case, $\hat{\textrm{x}}(\omega)$
acquires corresponding sidebands which, as illustrated in Fig.\ref{Fig1}(c), cause the two main peaks of  the shifted spectra 
${\hat{\textrm{X}}}^\pm(\omega)= \hat{\textrm{x}}(\omega+\omega_d)- \hat{\textrm{x}}(\omega-\omega_d)$ to interfere with each other's
sidebands. In this case, the minus sign in Eq.\ref{Yshif} (and the out-of-phase nature of the modulations) implies that one 
peak grows by constructive interference, while the other one diminishes. 

Full details are in \cite{SuppInfo}, but this can be understood from a simple argument.
For modest backaction (${\bar g}$ small), we
 can write ${\hat{\textrm{X}}}^\pm(\omega)$
in the form:
\begin{eqnarray}
 & {\hat{\textrm{X}}}^\pm(\omega) &\approx  \sqrt{\Gamma_M}[ {\hat X}_{th}(\omega+\omega_d)-  {\hat X}_{th}(\omega-\omega_d)] + \nonumber \\
& {\bar g} {\hat{\textrm{Y}}}_{BA}(\omega)& -i \omega_2 \sqrt{\Gamma_M} {\hat{\textrm{X}}}_{\omega_2}(\omega) 
-i \omega_2 {\bar g} {\hat{\textrm{Y}}}^{(\omega_2)}_{BA}(\omega)
\label{Shiftspec}
\end{eqnarray}
where the ${\hat X}_{th}$ terms represent incoming thermal noises, ${\hat{\textrm{Y}}}_{BA}$ represents the back-action terms driven by imprecision noise.
 The last two terms  are  corrections to account for the modulation of$\omega_M$  ; the first comprises thermal effects, the second
 the corresponding back-action effects.

\begin{figure}[t]
{\includegraphics[width=2.5in]{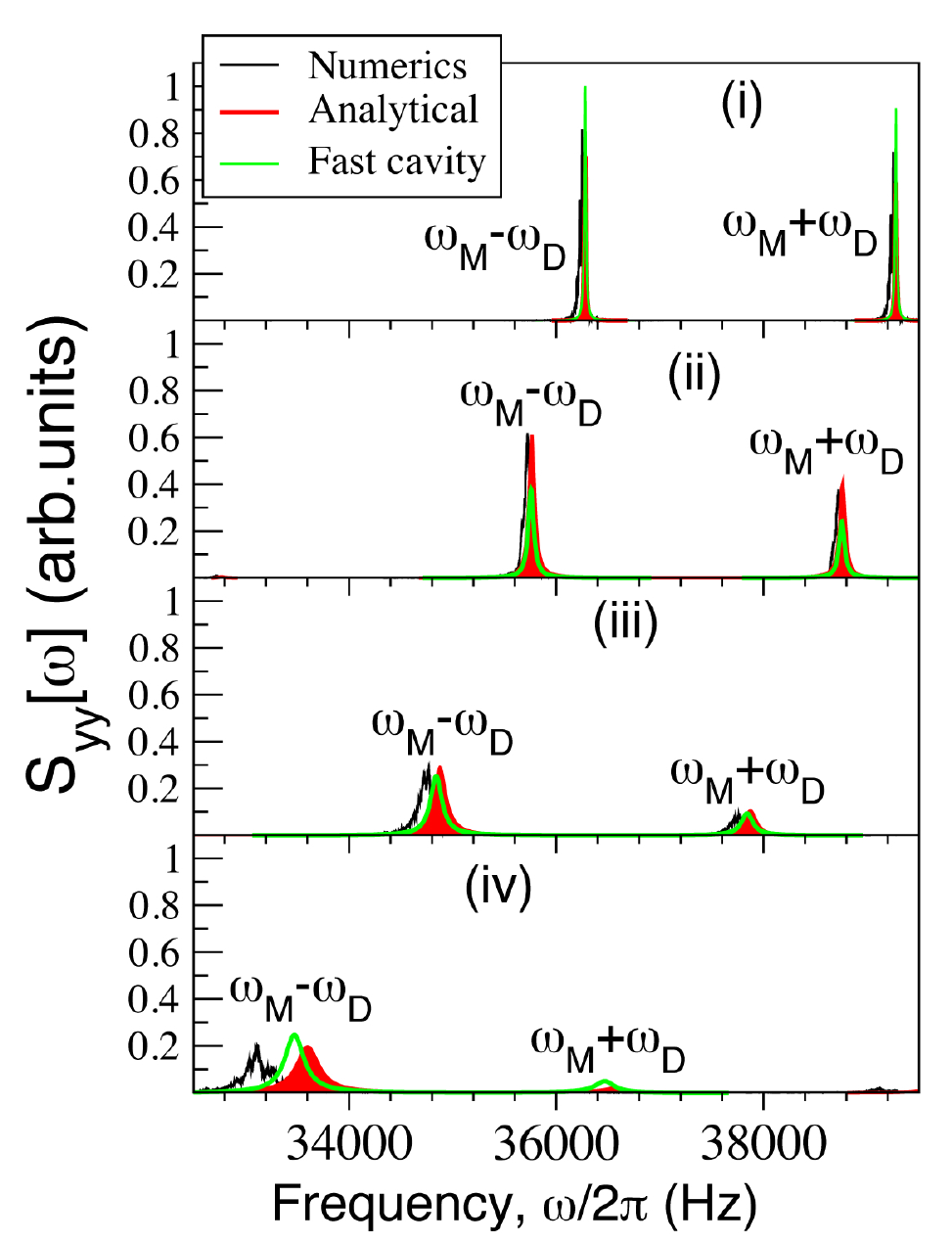}}
\caption{Comparison of the analytical split-sideband calculations with stochastic numerics and  fast cavity model,
with increasing ${\bar g},\omega_2$ for an optically trapped particle
 for thermal spectra, far from the quantum limit. Here, peak  
 heights scale with $\Gamma_M$ and $r$ is independent of $\Gamma_M$.
 In this regime, to obtain $S_{yy}(\omega)$ in units of Hz$^{-1}$, for arbitrary $\Gamma_M$, graphs should  be scaled 
as $S_{yy}(\omega) \times \Gamma_M/0.8$; in turn, for the optically trapped nanoparticles 
in \cite{arxiv2016},  $\Gamma_M \simeq 0.2 \times 10^4 P$, where the gas pressure ranges from $P=1-10^{-8}$ mbar.
 $\kappa/2=130\times 2\pi$ kHz,
$\Delta \simeq -75\times 2\pi$ kHz. Parameters are far from the sideband-resolved limit, so the fast-cavity model also gives reasonable results.
$N=100, 200, 300, 400$ in Eq.\ref{levitated} hence (i) $\omega_2/2\omega_d=0.05$, ${\bar g}=8500 \textrm{s}^{-1}$
 (ii) $\omega_2/2\omega_d=0.2$, ${\bar g}=17000  \textrm{s}^{-1}$
(iii) $\omega_2/2\omega_d=0.5$, ${\bar g}=25000  \textrm{s}^{-1}$,
(iv) $\omega_2/2\omega_d=0.9$, ${\bar g}=33000  \textrm{s}^{-1}$.}
\label{Fig2}
\end{figure}

 For $\omega_2 =0$ and neglecting backaction, 
 the shifted spectra arise mainly from  incoming thermal noises
$ {\hat X}_{th}(\omega)=\chi_M(\omega) {\hat{\textrm b}}_{in} + \chi_M^*(-\omega)  {\hat{\textrm b}}_{in}^\dagger$
weighted by the mechanical susceptibility $\chi_M(\omega) = \left [ -i (\omega - \omega_M) + \frac{\Gamma_M}{2} \right ]^{-1}$.
The anti-Stokes sideband for example, is primarily due to the weighted thermal noise operators 
$ \chi_M(\omega\pm \omega_d) {\hat{\textrm b}}_{in}(\omega \pm \omega_d)$. The susceptibilities $ |\chi_M(\omega\pm \omega_d)|$ are sharply
peaked at frequencies $\omega=\omega_M \mp \omega_d$ (since $\Gamma_M$ is small), yielding the ``twin peaks'' structure
since the ratio of the twin peak weights $r=|\chi_M(\omega - \omega_d)|^2/|\chi_M(\omega + \omega_d)|^2=1$.

The main effect of $\omega_2$ is to introduce the extra correction
from the ${\hat{\textrm{X}}}_{\omega_2}$ term which means replacing the thermal  weights:
\begin{eqnarray}
 \chi_M(\omega \pm\omega_d) &\to& \chi_M(\omega \pm \omega_d)[1-i \omega_2 \chi_M(\omega \mp \omega_d)]\nonumber\\
\label{noisea}
\end{eqnarray}
Evaluating the corrections (the terms in square brackets) near the frequency peaks of the noise, we find they are
$\approx (2\omega_d \pm \omega_2)/2\omega_d$ so the ratio of peaks in the PSD would be:
\begin{equation}
r \approx (2\omega_d - \omega_2)^2/(2\omega_d +\omega_2)^2,
\label{ratio}
\end{equation}
predicting a full cancellation for $2\omega_d \sim \omega_2$.  

For the standard optomechanical equations, Eq.\ref{Yspec}  and its ${\hat{\textrm{x}}}(\omega)$ equivalent are solved to obtain 
$S_{XX}(\omega)$ and $S_{yy}(\omega)$ or $S_{y_{out}y_{out}}(\omega)$ in closed form. 

However, for the modulated spectra this is not possible:
${\hat{\textrm{y}}}(\omega)$ depends on shifted ${\hat{\textrm{x}}}(\omega)$ spectra; and the $\omega_2$ modulation
couples ${\hat{\textrm{x}}}(\omega)$ spectra to the displacement spectra ${\hat{\textrm{x}}}(\omega\pm 2\omega_d)$.
Eqs.\ref{Yshif} and \ref{Shiftspec} are instead solved iteratively, assuming $\bar{g},\omega_2 \ll  \kappa,\omega_M$ 
and retaining terms up to cubic order in $\bar{g},\omega_2$ (see \cite{SuppInfo}). In the thermal regime, the resulting equations 
are tested against a set of numerical stochastic equations and another model, used to simulate optically trapped particles \cite{PRL2015,arxiv2016},
as seen in Fig.\ref{Fig2}.  

\begin{figure}[t]
{\includegraphics[width=3.0in]{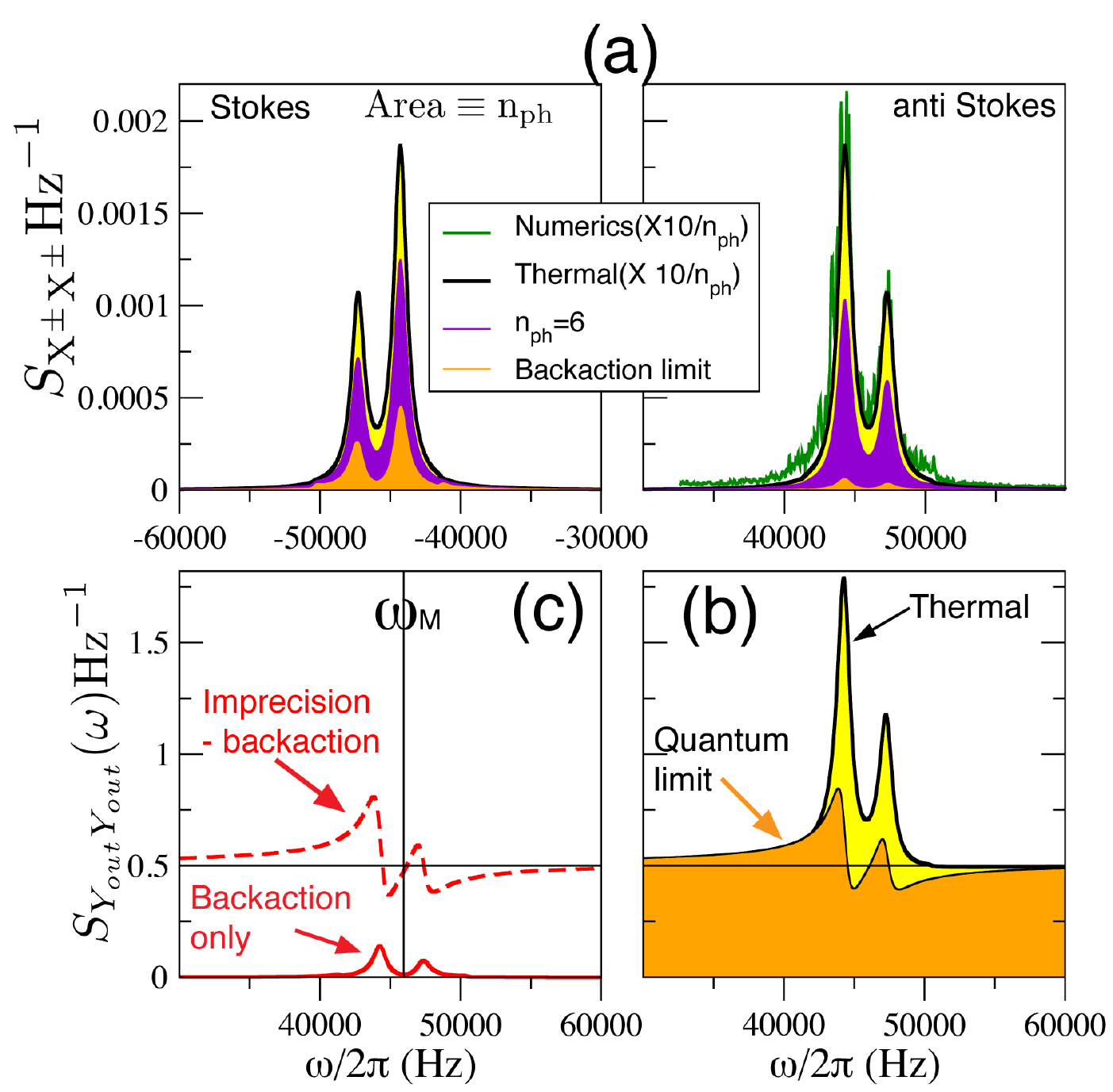}}
\caption{Comparison between the thermal spectrum and the quantum limit, using the analytical solutions
with increasing $g,\omega_2$  in the sideband-resolved limit, which can yield ground state cooling
at sufficiently low pressures. {\bf (a)} Shows $S_{{\textrm{X}}^\pm {\textrm{X}}^\pm}(\omega)$ for Stokes and anti-Stokes sidebands, as 
$\Gamma_{M} \to 0$ while the optomechanical cooling rate $\Gamma_{opt}$ in each graph remains fixed. The individual sideband
shapes are unchanged, but Stokes/anti-Stokes asymmetry develops. The symmetric classical spectra  are scaled to a height of 1
corresponding to $\Gamma_M=10^{-4}$s$^{-1}$. $\omega_2/2\omega_d=0.24$, for $2\omega_d=3 \times 2\pi$ KHz; $g=18,500$s$^{-1}$,
${\bar \omega}_M=46\times 2\pi$ kHz, $\kappa/2=26\times 2\pi$ kHz, $\Delta \simeq -\omega_M$.
 {\bf (b)}  The same solutions in (b) are now added to incoming  imprecision noise to obtain output spectra
$S_{Y_{out}Y_{out}}(\omega)$.  At high phonon occupancies, the shape is unchanged. As $n_{ph} \to n_{BA}$, the 
ratio above the quantum imprecision floor alters significantly.{\bf (c)} Shows individual contributions to the PSD; the pure backaction term has 
the same shape as the thermal split sidebands; its interference with incoming imprecision noise lowers the height of the
$\omega_M+\omega_d$ sideband.}
\label{Fig3}
\end{figure}

We then take $\Gamma_M \to 0$ which for cooling parameters (red-detuned light) takes
the system down to the quantum backaction limit,
where the heating is limited by quantum shot noise,   $n_{ph} \equiv n_{BA} \approx (\frac{\kappa}{4\omega_M})^2$ \cite{Peterson}.
When we calculate $S_{{\textrm{X}}^\pm {\textrm{X}}^\pm}$ (the PSD for ${\hat{\textrm{X}}}^\pm(\omega)$), we find that it differs very little from the thermal spectrum.
This indicates that even for $\Gamma_M=0$, a regime where the oscillator motion is completely driven by the incoming optical imprecision 
noise (quantum or classical in fact: this is true for a finite photon temperature), the shape, and $r$ for $S_{{\textrm{X}}^\pm {\textrm{X}}^\pm}$ unchanged as shown in Fig.\ref{Fig3}(a).

{\em However}, for $S_{y_{out}y_{out}}(\omega)$, this is not the case: when the same solution used for Fig.\ref{Fig3}(a) is added and interfered
with the incoming (imprecision) optical shot noise, the
 sideband shape is unchanged for the thermal regime but changes significantly in the quantum back-action limit.

The underlying reason for this change can be understood as follows: the total back-action in Eq.\ref{Shiftspec},
 ${\bar g}{\hat{\textrm{Y}}}_{T}(\omega)=  {\bar g} [{\hat{\textrm{Y}}}_{BA}(\omega)  -i \omega_2  {\hat{\textrm{Y}}}^{(\omega_2)}_{BA}(\omega)]$, which by itself
 still yields a ratio of $r$, develops correlations with the incoming imprecision terms ${\hat{\textrm{Y}}}_{imp}(\omega)={ \hat a}_{in} +{ \hat a}^\dagger_{in}-\sqrt{\kappa} \hat{Y}_{th}$. 
The key difference seen between $S_{{\textrm{X}}^\pm {\textrm{X}}^\pm}$ and  $S_{y_{out}y_{out}}(\omega)$ in the quantum limit, arise because:
\begin{equation}
\langle | {\bar g}{\hat{\textrm{Y}}}_{T}(\omega)|^2\rangle \neq \langle | \frac{{\hat{\textrm{Y}}}_{imp}(\omega)}{\sqrt{\kappa}}-  {\bar g}{\hat{\textrm{Y}}}_{T}(\omega) |^2\rangle
\label{Ymod}
\end{equation}
The above two terms are contrasted in Fig.\ref{Fig3}(c).
Ponderomotive squeezing originates from such correlations \cite{Safavi2013,Purdy2013,Pontin}
between backaction and incoming noise and, in the standard optomechanical case, it leads to a Fano-like line experimental
 profile \cite{Safavi2013,Purdy2013,Pontin}
and (an often small) dip where the output light spectrum lies below the imprecision floor.

However, in the present case, the height of the $\omega_M+\omega_d$ peak is lowered as it overlaps with a ponderomotive squeezing ``dip''
of the stronger peak as seen in  Fig.\ref{Fig3}(b), leading to a change in $r$: the sideband structure is more strikingly reshaped and the  $\Gamma_M$ invariance of $r$  is lost. Although ponderomotive squeezing does not require a ground state oscillator, for quantum shot-noise limited spectra,
 a change in $r$  only becomes appreciable if  $n_{ph} \to n_{BA}$, leading to a noticeable decrease in height of the $\omega_M+\omega_d$ peak above the
imprecision noise level.

\noindent {\em Stochastic numerical model} As outlined in \cite{PRL2015,arxiv2016},  a nanoparticle in a hybrid electrical-optical trap experiences a
dipole force potential $V(x)= -\hbar A |a(t)|^2\cos^2(k x)$ from the optical standing wave of a cavity (with axis along $x$). 
In \cite{arxiv2016}, the depth of the potential $A=26 \times 2\pi$ KHz, while the cavity photon number  $|a(t)|^2$ fluctuates
about mean value of $\approx 10^9-10^{10}$ photons; $k=2\pi/\lambda$ with $\lambda=1064$nm.
The particle becomes trapped in a given optical well $N$,
 with anti-node (potential minimum) at $x=X_N$ where $kX_n=2\pi N$.
 It experiences also an additional oscillating harmonic potential  $V^{\textrm{AC}}(x,t) = \frac{1}{2} m \omega^2_{\textrm{T}} (x+x_N)^2\cos (\omega_{\textrm{d}} t)$ 
from an ion trap.  We test our model by comparing with solutions of the equations of motion in these combined potentials, including also damping for the cavity ($\kappa$) and for mechanical degrees of freedom $\Gamma_M$) as well as stochastic Gaussian noise to allow for gas collisions and shot noise.
This represents a stringent test of our analytical noise model since, in the numerics, ${\bar g}, \omega_M$ and $\omega_2$ are not even input parameters: they are themselves emergent properties of the numerical simulations. For $\langle  |a(t)|^2\rangle \equiv \alpha$, we find:

\begin{eqnarray}
2kx_0(t) &\approx& -\frac{\omega^2_{\textrm{T}}}{\omega_{\textrm{M}}^2}(2kx_N)\sin(\omega_{\textrm{d}}t) \equiv X_d \sin(\omega_{\textrm{d}}t)  \nonumber\\
m\omega_{\textrm{M}}^2 &=&  2\hbar k^2 A|\bar{\alpha}|^2 \cos(2kx_0)\nonumber\\
2{\bar g} &=& k A \bar{\alpha} \sin(2kx_0).
\label{levitated}
\end{eqnarray}

hence the equilibrium point of the oscillations $x_0(t)$ oscillates as $\sin(\omega_{\textrm{d}}t)$, leading to modulated $\omega_M,{\bar g}$.

The fast mechanical motion $x_M(t)\simeq X_M \cos{\Phi_M(t)}$ where $X_M$ is the variance of the thermal motion,
 the phase being $\Phi_M(t)=\int \omega_M(t) dt$. 

 For a fast cavity, we can assume the cavity field follows $x(t)$ with no delay; to simulate this we combine the slow $x_0(t)$ motion with the
fast mechanical motion into the ansatz
$x(t)= X_d \sin(\omega_{\textrm{d}}t)+X_M \cos(\overline{\omega}_M + \frac{\omega_2}{2\omega_d} \sin{2 \omega_dt})$.
  The Fourier transform of $\cos 2kx(t)$
using this ansatz, gives a reasonable approximation of the split-sideband spectrum, for a fast cavity. Fig.\ref{Fig2} shows that it yields 
reasonable agreement with numerics and analysis. More importantly, it describes also scattering of light out of the cavity (illustrated in 
inset of Fig\ref{Fig1}(d)) which illustrated suppression of the $\omega_M+\omega_d$ sideband. While not a full demonstration, this classical-regime data 
does demonstrate the coherent relative phase accumulation and interplay between the slow and fast motions; it indicates that 
in combination with  homodyne or heterodyne detection, split sideband asymmetries may be investigated experimentally
once quantum-limited regimes are attained.

{\em Conclusions}  Split sideband spectroscopy offers a promising new experimental signature; measurement of the ratio $r$ complements  
Stokes/antiStokes asymmetry and offers an alternative probe of ponderomotive squeezing.
For a system where  the back-action spectra is noticeably reshaped by interference with incoming
noise, a striking signature of the quantum limit potentially exists in a single sideband, since  $r$ is well defined and controllable.
 Conversely, the double-sidebands may offer an additional diagnostic of Stokes/antiStokes asymmetry
as there are two pairs of peaks to compare.

 \end{document}